# CUMULATIVE PROCESSES OUT OFF THE NUCLEUS FRAGMENTATION REGION

S.S. Shimanskiy[1]

*JINR, 141980, Dubna, Moscow region, Russia*

## Abstract

In this report the first results of cumulative particle production in the new kinematical domain will be discussed. The produced particles have momenta more than 2 GeV/c in the rest frame of nuclear targets. Such studies can significantly reduce the corrections associated with the processes of the initial (ISI) and final states (FSI) interactions. We have made some conclusions from this data.

## Introduction

During last forty years were drastic changes in our knowledge about states of nuclear matter for different temperature and density. Fig.1 from [1]. shows the phase diagram (T(temperature)-$\mu_B$(baryon number density)) of nuclear matter as a function of time. The solid lines indicate a first-order phase transition where could be jumps of physical parameters. The dashed line is a crossover domain of where there are no jumps. The nowadays (t~2000) diagram shows that we deal with not only Quark - Gluon Plasma (QGP) but some new state of nuclear matter - the cold high density state for which dozen models with quarkyonic, diquarks and other exotic configurations have been proposed [2]. Why this domain (low temperature and high density) of the phase diagram is so interesting for investigation?

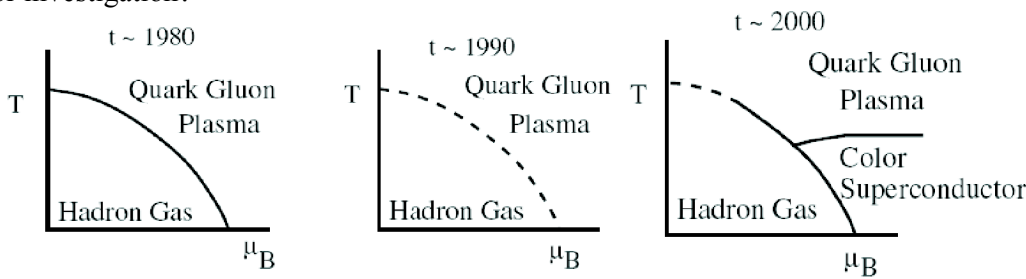

Fig.1. The phase diagram of nuclear matter evolution with time.

The cold dense baryonic(nuclear) matter (CDNM) exists as a core part of massive stars. Properties of this state determine physical characteristics (mass, radius and etc.) and scenarios of evolution of massive stars. The large number of models proposed to describe the cold dense baryonic matter means that the general (mass, radius and etc.) physical properties of massive stars do not allow for a rigorous selection of the models. That is why it is so important to recognize whether we have the opportunity to explore this domain of the phase diagram in laboratory experiments. The CDNM must have a density at least of 5-10 times greater the nuclear density. To obtain this state of the nuclear matter, we need to push in to the volume of the nucleon a few nucleons. In this case, we are likely going to have deal not with the hadronic but the quark-gluon degrees of freedom. In other words we need to find multinucleon (multiquark) configurations in the ordinary nucleus.

Therefore high energy accelerators with nuclear beams are needed if CDNM does not exist in the initial state and can only be created as a result of a nucleus-nucleus collisions.

---

[1] E–mail: Stepan.Shimanskiy@jinr.ru

But the energy of the accelerators should not be too high as the energy increases the nuclear stopping power and the hard scattering cross sections decrease rapidly. Besides accelerators with appropriate energy of nuclear beams we need to know signatures about CDNM formation. Moreover in the heavy ions collisions can be formed hot dense baryonic matter which needs to separate from the signals of CDNM. That's why for these studies more preferable to use the light ion collisions.

If CDNM exists as a quantum component of the wave function of ordinary nuclear matter (i.e. exists in the initial state) it can be detected not only in the nucleus-nucleus collisions but in collisions with nucleus different particles as neutrinos, photons, leptons and etc.

How can we register CDNM or what signals will talk about CDNM formation? The existence of the CDNM means that we must observe the interaction with or formation of the multinucleon (multiquark) configurations. There are two ways to find and to examine CDNM. The first, we can study the excitation spectra of CDNM and the second opportunity to investigate scattering different particles in a similar way as has done by Rutherford.

Since we are interested multinucleon (multiquark) configurations to minimize background signals we need to examine the kinematic domain where several nucleons participate in the studied processes. The production or scattering of particles in the kinematical region where requires the participation of several nucleons were called cumulative processes [3].

1. **Cumulative processes**

Cumulative processes intensively began to be studied after 1971 when A.M.Baldin has published an article [3] in which he predicted that using nucleus beams can produce mesons with energy much more than per nucleon in the beam. Since that time, detailed studies of the inclusive spectra of cumulative particles were conducted at JINR (Dubna) and ITEP (Moscow). Some later, other institutions have joined to these investigations. The data and the theoretical models which were developed to describe the data observed in review articles [4].

The bulk of experimental data of cumulative processes have been obtained by the inclusive setups in fragmentation regions of beams or targets. The very important features have been discovered. The slope $T_0$ of the cumulative particle spectra ( if inclusive cross section is approximated as $f = E\frac{d^3\sigma}{dp^3} = C \cdot exp(-\frac{T}{T_0})$ ) is independent on type and energy of incident particle and A (A is a mass number of nucleus) of fragmentation nucleus Fig.2 from [5]. From Fig.2 we can make conclusion that the cumulative particles haven't produced during compression but should be produced by the CDNM configuration which exist in ordinary nuclear matter before collisions as some fluctuation of nuclear density.

Isotopic symmetry was discovered - the ratios of cumulative hadrons (neutron to proton [5] and $\pi^+$ to $\pi^-$ [6]) for the nonsymmetric heavy nuclei approaches unity when the hadron energy increases. This distinguishing feature says that the CDNM is an isotopic singlet state mainly with identical number of u and d quarks. We need to stress that the same effect has been observed at JLAB experiments [7] more lately too.

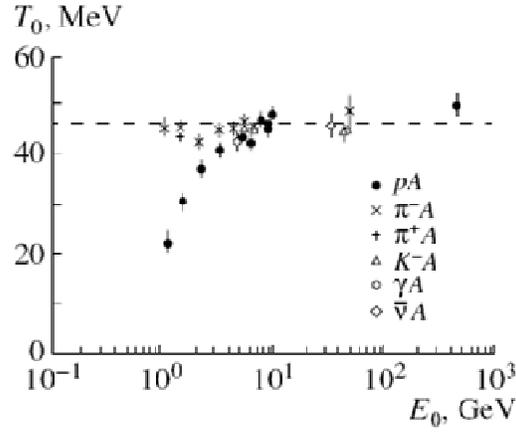

Fig.2. The slope parameters $T_0$ for inclusive proton spectra [5] for different projectile $p, \pi^{\pm}, K-, \gamma, \bar{\nu}$ with various energies $E_0$ (the production angle is $120^o$ in the laboratory frame).

The Fig.3 shows that the inclusive spectra of cumulative particles [5] as a function of the light cone variable $\alpha = \frac{(E - p_L)}{m_N}$ are similar at $\alpha > 1$ domain and the cross sections decrease with increasing number of constituent sea quarks (non u or d quarks).

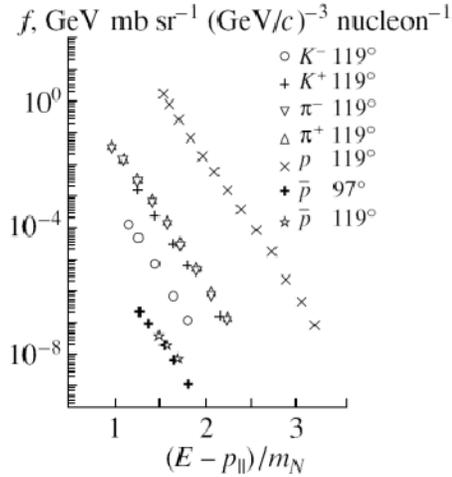

Fig.3. The spectra of the protons, $\pi^{\pm}$ mesons, $K^{\pm}$ mesons, and antiprotons from p-Cu interaction of 10 GeV proton beam (registration angle is 119° in the laboratory frame) versus light-cone variable $(E - p_\_)/m_N$.

This data (Fig.3) shows that the sea components at CDNM are suppressed. It means that inside ordinary nuclear matter exist blobs of CDNM which consist of u and d quarks mainly. Moreover, because at cumulative particles production are involved few of nucleons the strong A-dependence of cross sections must be and have been observed [4]. The studies of inclusive spectra were allowed to determine the probabilities of existence multinucleon (multiquark) configurations for different nuclei [8].

As was mentioned above, the second method to find CDNM is a scattering of a particles (in particular leptons) on nuclear targets to define the mass $x$ ($x$ is a Bjorken variable) on which took place scattering. The deep inelastic scattering (DIS) experiments at $x > 1$ domain are performed with lepton beams since 80-th years of the last century [9]. Detail JLab investigation with electron beams [10] gave possibility to derive the probabilities of multinucleon (multiquark) configurations in nuclei too. Comparison of the probabilities of multinucleon (multiquark) configurations shows that as the data from the spectra study well as the DIS data give similar values. So it can assume that we have serious evidence the existence in nuclei the droplets of CDNM. If it so then the next step should be a detailed study of CDNM properties.

Most of the cumulative data have been obtained in the nuclear fragmentation and the low transverse momenta ($P_T$) domain where there are significant corrections through the initial state (ISI) and the final state (FSI) interactions. To avoid these problems we need to go out of the fragmentation region (p > 1-2 GeV/c in the rest frame of nucleus) and to go in the high $P_T$ region ($P_T > 1$ GeV/c). In this kinematic region can expect a significant contribution from the direct knock out of multinucleon (multiquark) configurations.

The experiment E850/EVA at BNL [11] the first one began to explore this new kinematic region in the correlation measurements of ($p, 2pn$) reactions. Unfortunately, this experiment studied only signals of existence of the two nucleons correlation so-called Short Range Correlations (SRC) Fig.4a. The SRC model based on the possibility to exist in nuclei correlated configuration of the white point like nucleons with the large relative momenta [12]. It was concluded that 2N-SRC must be a major source of high-momentum nucleons in nuclei. Since nucleons have the large relative momenta (up to ~ 1 GeV/c) and therefore they are at small distances it is one of the possible forms of CDNM. In this experiment has not been investigated another possible forms of existence CDNM in the form of the multiquark bags. The models consider different quark bags back to model D.I. Blokhintsev [13] who considered the fluctuations of the nuclear density. Accepted name for this multinucleon (multiquark) bag is a "flucton". On the Fig.4b shows reaction with participation of the "flucton". In case of the "flucton" we should observe in the region of recoil particles not only protons (as for SRC) but also nuclear fragments and jets of unbinding nucleons (for example, system of few neutrons or protons).

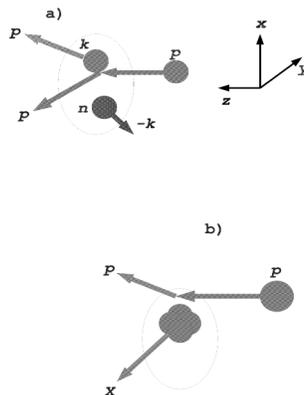

Fig.4. a) registration of the two **2N-SRC**; b) detection of the **"flucton"**.

In [14] was proposed to conduct a detailed study of the recoil particles to determine the kinematic regions where the "flucton" mechanism and where the SRC mechanism are dominated.

## 2. Cumulative processes far from nuclear fragmentation domain [16,17]

The high $P_T$ and far from nuclear fragmentation domain where cumulative processes have not yet been studied experimentally, is of particular interest. A theoretical analysis and experimental data said [15] that performed in revealed that, at $X_T \sim 1$, processes of interaction with multiquark (multinucleon) configurations must make a dominant contribution, the contribution of background rescattering processes being small. Here $X_T = 2P_T/\sqrt{s}$ is a fraction of the maximum possible transverse momentum, with $s$ being the invariant energy of nucleon–nucleon interaction.

The first experimental data in this new kinematical domain were obtained at a single-arm magnetic spectrometer SPIN (IHEP, Protvino) Fig.5 [16]. In 2009-2011 positive charge particle spectra have been measured only.

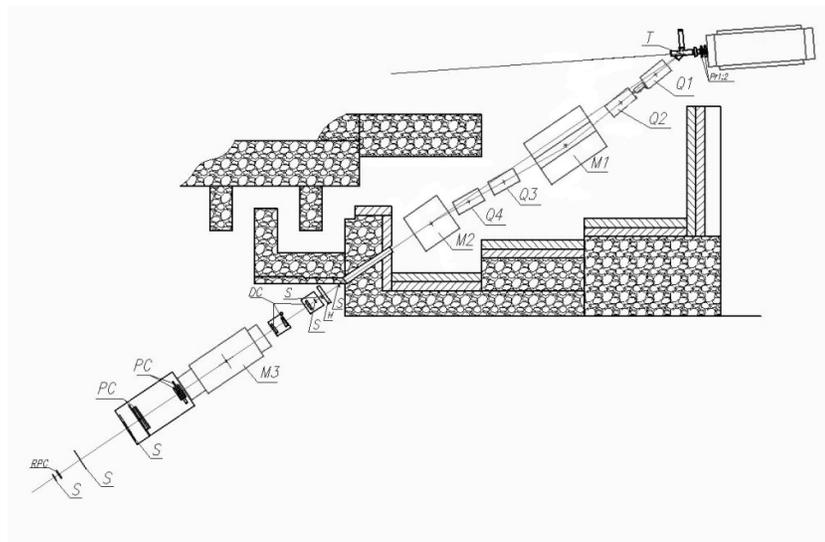

Fig. 5. Schematic diagram of the spectrometer SPIN: T - target; Q1, Q2, Q3, and Q4 – magnetic lenses; M1, M2 – dipole magnets selecting particles that leave the target at different angles; M3-analyzing magnet; S - set of scintillation trigger counters; PC, DC - wire chambers of the tracking system; H (hodoscope), and RPC (resistive plate chamber) – elements of the time of flight system.

Cumulative processes are characterized by a strong dependence on the atomic number [4]. In approximating the cross section by a power-law function that is proportional to $A^\alpha$, the value of $\alpha$ for cumulative particles may be greater than unity. The exponent $\alpha$ obtained from the tungsten-to-carbon cross-section ratio is given in Fig.6 as a function of the transverse momentum [16]. This figure shows that $\alpha$ grows with $P_T$, reaching a value of about 1.37 at $P_T \approx 3.5$ GeV/c. A strong $A$-dependence may be the result of interaction with CDNM (local interaction) or due to the rescattering of secondary hadrons on intranuclear nucleons.
+

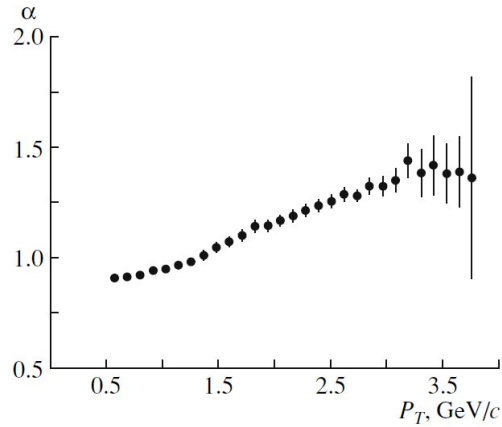

Fig. 6. Exponent in the power-law dependence of the cross section on the atomic number as a function of the transverse momentum [16].

Moreover, in [16] was shown that the spectra obtained in the subcumulative region deviate substantially from the respective predictions of the standard Monte Carlo generators UrQMD and HIJING, which purport to describe proton–nucleus and nucleus–nucleus interactions. In the cumulative domain these MC generators have no predictions. In 2012 spectra positive and negative particles have been measured [17]. The experimental results for $h^+/h^-$ - ration are presented at Fig.7.

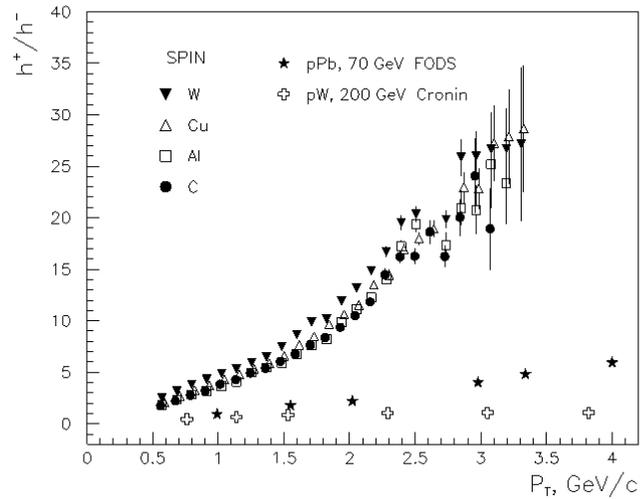

Fig.7. $h^+/h^-$ data of this experiment for different targets as a function of transverse momentum. The data on the yield ratios of protons and $\pi^-$ - mesons, $p/\pi^-$, for pPb at 70 GeV [18] and pW at 200 GeV [19] are shown for comparison. The two latter measurements were made at a registration angle of $90^0$ in the center of mass system for free pp -interactions.

For all four targets used in the experiment the ratio of the yields of positively charged ($h^+$) and negatively charged ($h^-$) particles grows fast with increasing transverse momentum. Fig.7 shows the $h^+/h^-$ - this experiment data as a function of $P_T$ and for comparison the $p/\pi^-$ data obtained at comparable transverse momenta in $p$-$Pb$ collisions at 70 GeV proton energy and a registration angle of 9° in the laboratory frame [18] as well as the data [19] obtained in $pW$ interactions at 200 GeV proton energy and a registration angle of 4.4° (in both experiments [18] and [19] an angle of 90° in the center of mass system for free nucleons was chosen).

It should also be noted that with increasing $P_T$ the $h^+/h^-$ data of this experiment become close for all targets. For illustration at Fig.8 shows comparison of the ratio of $h^+/h^-$ for tungsten and for aluminum. The absence of strong dependence of $h^+/h^-$ on atomic number at high $P_T$ can be interpreted as an indication of **a local mechanism of particle production** and small contribution of secondary interactions.

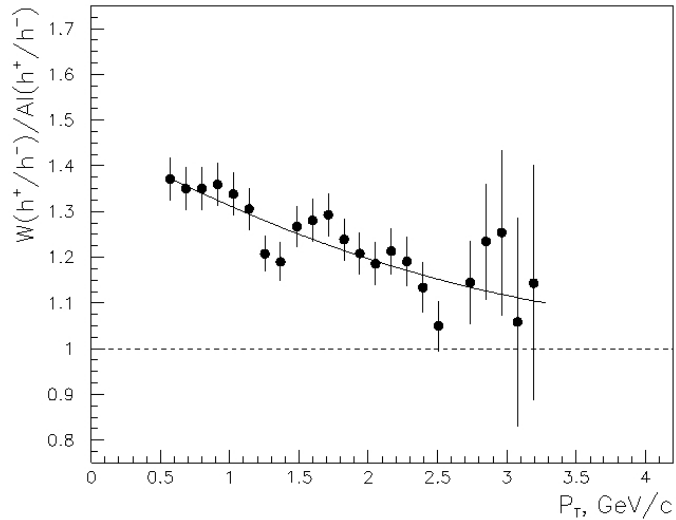

fig.8. Ratio of positively and negatively charged particle yields $h^+/h^-$ measured on W and Al nuclei is compared. The curve shows a polynomial approximation.

Cumulative processes are characterized by high four-momentum transfer. If a cumulative process is considered as a quasi-binary subprocess in which fractions $X_1$ and $X_2$ of four-momenta of the incident particle and the target respectively take part its can be described using these self-similarity variables. This approach is similar to that used in the parton model. The difference is that in this case the squared four-momentum is nonzero, it is equal to the squared fraction of the nucleus (hadron) mass participating in the subprocess. A similarity parameter $s_{min}$ was proposed by V.S.Stavinskiy [20] for description of cumulative processes and high-$P_T$ reactions; this parameter is based on the minimization of a functional comprised of fractions of four-momenta with the meaning of minimum squared sum of fractions of four-momenta of colliding objects necessary for production of the registered pair of particles. This approach makes it possible to determine unambiguously the fractions of four-momenta $X_1$ и $X_2$ corresponding to this minimum for any particular reaction. The analysis of a large set of inclusive data on particle production in the pre-cumulative and cumulative domains and in inclusive processes of low-$P_T$ subthreshold particle production [21] demonstrated ability to provide qualitative description of invariant particle production cross sections in the form of an incomplete self-similarity solution:

$$f = E\frac{d^3\sigma}{dp^3} = C_1 \cdot A_1^{\frac{1}{3}+\frac{X_1}{3}} \cdot A_2^{\frac{1}{3}+\frac{X_2}{3}} \cdot e^{-\frac{\Pi}{C_2}} \quad (1)$$

where $C_1$ and $C_2$ are constants, same for all reactions, $A_1$ and $A_2$ are the atomic masses of colliding nuclei, $\Pi = \frac{\sqrt{s_{min}}}{2m_N}$ is the similarity parameter where $s_{min}$ is the squared minimum energy for the calculated fractions of four-momenta necessary for producing the observed particle, and $m_N$ is the nucleon mass (in the case of nucleon and nuclear collisions). For proton-nucleus interaction $A_1 = 1$. It was shown in [21] that self-similarity solution (1) provides general regularities of the behavior of multiple processes, especially in description of A-dependences of inclusive cross sections. According to (1), the ratio of invariant cross sections multiplied by inverse A-dependence should be equal to unity for $pA$ interaction.

$$\frac{f_{(p+A_I)}}{f_{(p+A_{II})}} \times \left(\frac{A_I}{A_{II}}\right)^{-(\frac{1}{3}+\frac{X_2}{3})} = 1 \quad (2)$$

Here $A_I$ and $A_{II}$ are the atomic masses of the target nuclei. It is important to verify whether relation (2) is satisfied for a process similar to a cumulative one. Fig.9a shows the ratios of $\pi^-$ production cross sections multiplied by $A^{-(1+X_2)/3}$ for different nuclei. The lower axis of the figure shows the momentum, the upper, the quantity $X_2$.

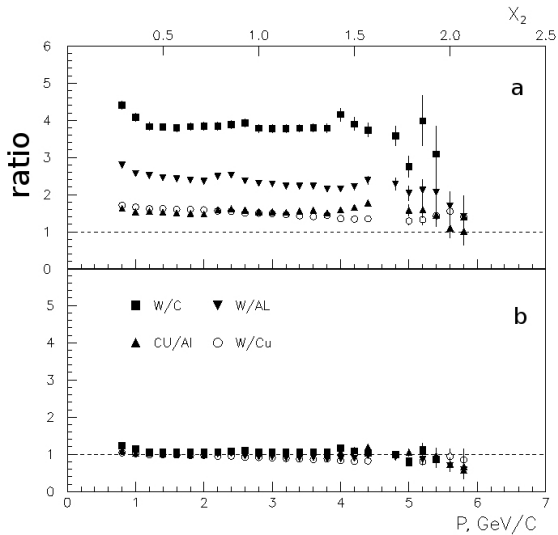

Fig.9. Ratio of cross sections of negative pion production on different nuclei multiplied by inverse A-dependence (see the text). The lower axis shows the momentum, the upper axis, $X_2$. (a) The ratios are obtained using the A-dependence in the form [21] $A^{(1+X_2)/3}$, (b) the ratios are obtained using the A-dependence in the form $A^{(2.45+X_2)/3}$.

It can be seen from this figure that all four ratios W/C, W/Al, W/Cu, and Cu/Al are close to a constant, however, the ratios themselves differ from each other strongly for different pairs of nuclei. The A-dependence of the form $A^{(1+X_2)/3}$ well describes the dynamic dependence of the cross sections on $X_2$, however, it fails to describe a stronger dependence on the mass number of the nucleus observed in this experiment. By using the A-dependence of the form $A^{(\alpha+X_2)/3}$ it was found that the presented negative pion production cross sections correspond to the parameter α=2.45±0.04. This is illustrated in Fig.9b this figure shows the ratios of the same cross sections as in Fig.9a, each multiplied by $A^{-(2.45+X_2)/3}$.

The data obtained in [16,17] confirm the main regularities observed in previous studies of cumulative processes which have been discussed in section **1**. An important new result of these studies that in the new kinematic domain it is possible to carry out a detailed study of CDNM properties.

**Conclusion and Future**

Which main results have been received in [16,17}? The momentum spectra of cumulative charged particles in the high-$P_T$ (up to 3.5 GeV/c) domain in *pA*-reactions were obtained for the first time. The strong dependence of the measured cross sections on the mass number of nuclei and the reasonable assumption of **the local character of the particle production** make it possible to interpret these data as an indication of a large contribution of the processes of interaction of the incident proton with multinucleon (multiquark) configurations or CDNM.

The observation of cumulative particles in the high $P_T$ and far from nuclear fragmentation domain makes it possible to plan next step the correlation studies of the properties of CDNM component of nuclear matter. In the correlation experiments when a cumulative particle can be used as a trigger, and registration of the accompany particles with it can provide detail information on the nature of CDNM. Currently under discussion the possibility of upgrading SPIN setup with the aim to create a second arm of the spectrometer for carrying out the correlation measurements.

The JINR-ITEP collaboration creates at VBLHEP JINR(Dubna) the MARUSYA-FLINT setup which will allow to perform the correlation measurements at the nuclotron with energy of a nucleus beams up to 5 A GeV [22].

This work is partly supported by RFBR grant № 14-02-00896.